\documentclass[a4paper,twocolumn,english,aps,prb,floatfix]{revtex4}
\usepackage[latin1]{inputenc}
\usepackage{amsmath}
\usepackage{babel}
\usepackage{graphics}
\usepackage{amssymb}
\usepackage{color}
\usepackage{bm}

\def\ave#1{\langle #1\rangle}

\newcommand{\up}{\uparrow}
\newcommand{\dn}{\downarrow} 
\newcommand{\iv}{\mathbf i}
\newcommand{\jv}{\mathbf j}

\begin{document}


\title{Kondo--attractive-Hubbard model for the ordering of local magnetic moments in superconductors} 

\author{Pedro R.\ \surname{Bertussi},$^1$ Andr\'e L.\ \surname{Malvezzi},$^2$ 
        Thereza \surname{Paiva},$^1$ 
        and 
        Raimundo R. \surname{dos Santos}$^1$} 

\affiliation{$^{1}$Instituto de F\'\i sica, 
                 Universidade Federal do Rio de Janeiro, 
                 Caixa Postal 68528,
                 21941-972 Rio de Janeiro RJ, 
                 Brazil\\
             $^2$Departamento\ de F\'\i sica,
                 Faculdade de Ci\^encias,
                 Universidade Estadual Paulista,
                 Caixa Postal 473,
                 17015-970 Bauru SP,
                 Brazil}
\begin{abstract}
We consider local magnetic moments coupled to conduction electrons with on-site attraction, in order to discuss the interplay between pairing and magnetic order. 
We probe the ground state properties of this model on a one-dimensional lattice through pair binding energies and several correlation functions, calculated by means of density-matrix renormalization group. 
A phase diagram is obtained (for fixed electron density 1/3), from which we infer that coexistence between magnetic order and superconductivity is robust, at the expense of a continuous distortion of the magnetic arrangement of the local moments, as evidenced by a strong dependence of the characteristic wave vector ${\bf k}^*$ with the coupling constants. This allows us to understand some trends of the coexistence, such as the influence of the rare earth on ${\bf k}^*$, as observed experimentally in the borocarbides.
\end{abstract}

\date{\today}


\maketitle  

The advent of high temperature cuprate superconductors in the late 1980's singled out the interplay between superconductivity and magnetic order. 
This was followed by the discovery of robust coexistence between some degree of magnetic order and superconductivity in ternary and quaternary rare earth compounds,\cite{Cava94,Muller01,Baba08} as well as in heavy fermion matter.\cite{Saxena00,Flouquet06} 
Very recently, a new class of FeAs-based superconductors has attracted a lot of attention due to their (moderately) high critical temperature, and new experimental evidence has been gathered indicating that superconductivity coexists with a spin-density wave state in some members of the ferropnictide family.\cite{Chen09,Xiao09} 
In spite of these experimental advances, microscopic modelling of coexistence between magnetic and superconducting orderings is still in its infancy, and considerable insight should be gained by investigating the competition between these two opposing tendencies. 

With this in mind, here we focus on a specific class of materials showing this coexistence, namely, the borocarbides, in which the rare earth element provides local moments (through their $f$-electrons), while superconductivity arises from phonon-mediated pairing of conduction electrons: singlet superconductors with either antiferromagnetic or modified ferromagnetic (i.e., spiral or domain-like) arrangements have been observed experimentally.\cite{Muller01}
We assume that pairing of conduction electrons can be described by the attractive Hubbard model,\cite{Micnas90} and that they are coupled to local moments through a Kondo-like term.\cite{Tsunetsugu97,Garcia04} The Hamiltonian then reads
\begin{eqnarray}
\label{Ham} 
{\cal H}&=&-t\sum_{\langle\iv,\, \jv\rangle,\sigma}
\left(c_{\iv\sigma}^\dagger c_{\jv\sigma}^{\phantom{\dagger}}+\text{H.c.}\right)  
- U \sum_\iv n_{\iv\uparrow}n_{\iv\downarrow} 
\nonumber\\
&&+ J\sum_{\iv}{\bf S}_\iv \cdot \bm{\sigma}_\iv\;,
\end{eqnarray} 
where, in standard notation, the sums run over lattice sites, with $\langle\iv,\, \jv\rangle$ denoting nearest neighbor sites, $t$ sets the energy scale (we set $t=1$ from now on), $U>0$ is the attraction strength, and $J$ is taken positive, thus favoring an antiferromagnetic coupling between the conduction electron spin $\bm{\sigma}_i\equiv \sum_{\alpha,\beta=\pm}c_{i\alpha}^\dagger\bm{\sigma}_{\alpha\beta} c_{i\beta}^{\phantom{\dagger}}$ ($\bm{\sigma}_{\alpha\beta}$ denotes the Pauli matrix elements) and the localized spin ${\bf S}_i$; for simplicity, we take $S=1/2$. 
The two competing tendencies are clear: as $J$ increases, the Kondo-like coupling drives the conduction electrons to form singlets with the local moments, at the expense of breaking the pairs.   

In order to determine the properties of this model in an unbiased way, we consider a one-dimensional lattice and resort to the density matrix renormalization group (DMRG) \cite{White92,White93,Malvezzi03,Hallberg06} to obtain the ground-state $| \psi_0 \rangle $ and energy, $E_0(N_e)$, where $N_e$ is the number of electrons. 
The aspects of competition which will be highlighted here do not depend qualitatively on the fact that one-dimensional 
ordering can only be quasi--long-ranged. 

The Hamiltonian (\ref{Ham}) is investigated on lattices with $N_s$ sites (hence $8^{N_s}$ states in the full Hilbert space) and open boundary conditions. For the sake of brevity, we restrict ourselves to 
the density $n=N_e/N_s=1/3$, though we have also examined other densities, and the results are qualitatively the same, apart from half filling.
Lattice sizes up to 60 sites were used, and truncation errors in the DMRG procedure were kept around $10^{-5}$ or smaller. 

The superconducting state is probed with the aid of the pair binding energy, 
\begin{equation}
E_B=2E_0(N_e+1)-E_0(N_e+2)-E_0(N_e)\,,
\label{EB}
\end{equation}
a positive value of which indicates that the state is superconducting, and of the $s$-wave pairing correlation function,
\begin{equation}
P_s(r)=\langle c_{i\dn}^\dagger c_{i\up}^\dagger c_{i+r\up}^{\phantom{\dagger}} c_{i+r\dn}^{\phantom{\dagger}} +\text{H.c.} \rangle.
\label{Ps}
\end{equation}
The magnetic properties of the local moments are probed with the
real-space correlation functions, 
\begin{equation}
S^{\mu\mu}(r)=\langle  S^{\mu}_i S^{\mu}_{i+r} \rangle,\ \mu=x,y,z\;,
\label{Smumu}
\end{equation}
and their corresponding 
structure factors, 
\begin{equation}
\tilde{S}^{\mu\mu}(k)= \frac{1}{N_s}\sum_{i,j}{\rm e}^{ik(i-j)} \langle  S^{\mu}_i S^{\mu}_{j} \rangle,\ \mu=x,y,z\;,
\label{Smuq}
\end{equation}
as well as their sums,
\begin{equation}
\tilde{S}(k)= \sum_{\mu=x,y,z} \tilde{S}^{\mu\mu}(k)\;,
\label{sigmaq}
\end{equation}
where $\langle  \ldots \rangle\equiv \langle \psi_0|  \ldots|\psi_0\rangle$.

\begin{figure}
{\centering\resizebox*{8.8cm}{!}{\includegraphics*{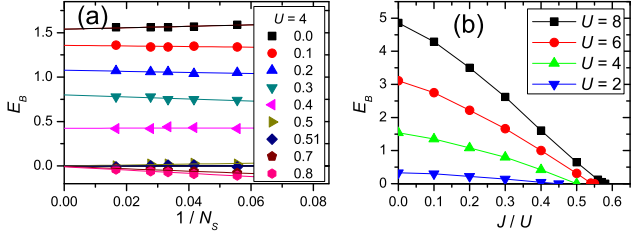}}}
\caption{(Color online) (a) Pair binding energy [Eq.\ (\ref{EB})] as a function of the inverse lattice size, for $U=4$, for the values of $J/U$ labelling the curves; data are taken for $N_s=18$, 24, 30, 36 and 60 sites.
(b) Extrapolated (to $N_s\to\infty$) pair binding energy as a function of $J/U$ for the values of $U$ labelling the curves; the error bars are smaller than the data points.}
\label{E_B} 
\end{figure}

\begin{figure}
{\centering\resizebox*{8.8cm}{!}{\includegraphics*{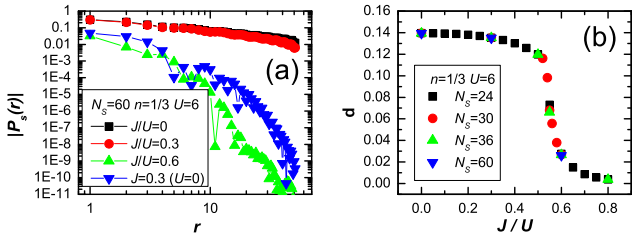}}}
\caption{(Color online) (a) Spatial dependence of the pairing correlation function [Eq.\ (\ref{Ps})], in a double-logarithmic scale, for a lattice with 60 sites, and $U=6$ in all cases but one, for which $U=0$;
(b) average double occupancy (see text) as a function of $J/U$, for different system sizes.}
\label{Ps60} 
\end{figure}

For given values of $U$ and $J$, we calculate the binding energy for $N_s=18$, 24, 30, 36 and 60 sites, which, when plotted as functions of $1/N_s$, allows for smooth extrapolations to $N_s\to\infty$; the result is shown in Fig.\ \ref{E_B}(a).  
For each $U$, the binding energy decreases as the Kondo coupling increases, and vanishes at a critical value, $(J/U)_c$, signalling the breakdown of superconductivity; see Fig.\ \ref{E_B}(b), which also shows that $(J/U)_c$ decreases as $U$ decreases. 
It is important to check this against the behavior of other quantities, for consistency. 
Figure \ref{Ps60}(a) shows the spatial decay of the pairing correlation function, Eq.\ (\ref{Ps}), 
for $U=6$ and, for comparison, one case for $U=0$. 
(In order to discard edge effects due to open boundaries, in our plots of spatially dependent quantities, we place the origin at $i=5$.)  
For $J=0$, the squares in Fig.\ \ref{Ps60}(a) reproduce the behavior of $P_s$ for the attractive Hubbard model, which definitely displays quasi--long-range superconducting order;\cite{Marsiglio97,Guerrero00,Salwen04} by contrast, when $U=0$ [down triangles in Fig.\ \ref{Ps60}(a)] the system is certainly not superconducting, and the correlation function near the chain edge is at least seven orders of magnitude smaller than in the previous case.
In-between these extreme cases, e.g., when $J/U=0.3$, the pairing correlation function (circles) can hardly be distinguished from that for $J/U=0$, 
while for $J/U=0.6$, the correlations decay as fast as when $U=0$. 
This change in spatial decay rate is accompanied by a significant drop in the average site double occupancy, 
$d\equiv(1/N_s)\sum_{i}\langle n_{i\uparrow} n_{i\downarrow}\rangle$, as it can be seen from Fig.\ \ref{Ps60}(b).
These predictions therefore agree with a superconducting transition taking place at $(J/U)_c\simeq0.55$ for $U=6$, as determined from Fig.\ \ref{E_B}(a). 
 
\begin{figure}
{\centering\resizebox*{8.8cm}{!}{\includegraphics*{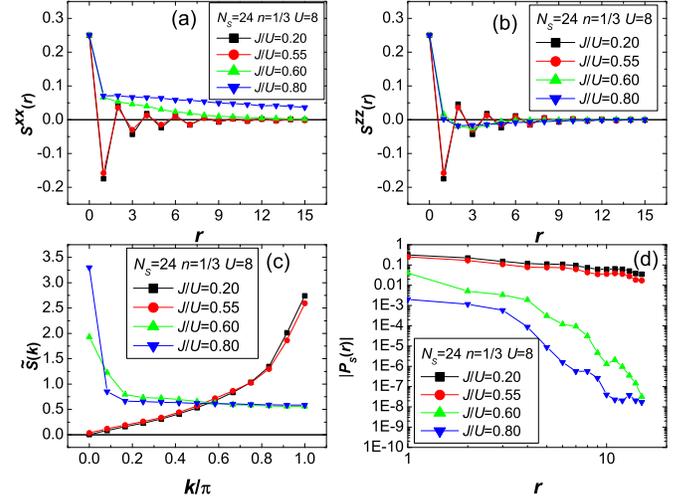}}}
\caption{(Color online) Results for a lattice with 24 sites, $U=8$, and for different values of $J/U$: (a) Spatial dependence of the local-moment $xx$-correlation function [Eq.\ (\ref{Smumu})] (b) same for the $zz$-correlation function; (c) wavevector dependence of the structure factor; 
(d) log-log plot of the spatial dependence of the pairing correlation function [Eq.\ (\ref{Ps})]}
\label{dataU8} 
\end{figure}

In Fig.\ \ref{dataU8} we correlate the presence of local-moment magnetism and superconductivity, in the regime of strong attraction ($U=8$).
For small values of $J/U$, the local-moment correlations are isotropic, i.e., $\ave{S_i^xS_{i+r}^x}=\ave{S_i^zS_{i+r}^z}$, and a spin-density wave (SDW) with period 2 (peak of $S(k)$ at $k=\pi$) is formed; see plots for $J/U=0.2$ and 0.55 in Figs.\ \ref{dataU8}(a)-(c). 
Fig.\ \ref{dataU8}(d) shows that in this same regime of $J/U$, pairing correlations are slowly decaying, consistent with a superconducting (SC) state. 
As $J/U$ increases beyond 0.6, one enters a ferromagnetic (FM) 
region characterized by $\ave{S_i^xS_{i+r}^x}$ displaying ferromagnetic behavior, while $\ave{S_i^zS_{i+r}^z}$ only displays short ranged SDW correlations; this picture is supported by the magnetic structure factor now showing a peak at $k=0$, as it can be seen from Fig. \ref{dataU8}(c). 
This change in magnetic behavior is accompanied by a drastic change in pairing correlations [Fig. \ref{dataU8}(d)]: they become  non-superconducting in the FM region. 
Again, the superconducting transition is consistent with the value $(J/U)_c\simeq 0.58$, extracted from Fig.\ \ref{E_B}(b).

\begin{figure}
{\centering\resizebox*{8.8cm}{!}{\includegraphics*{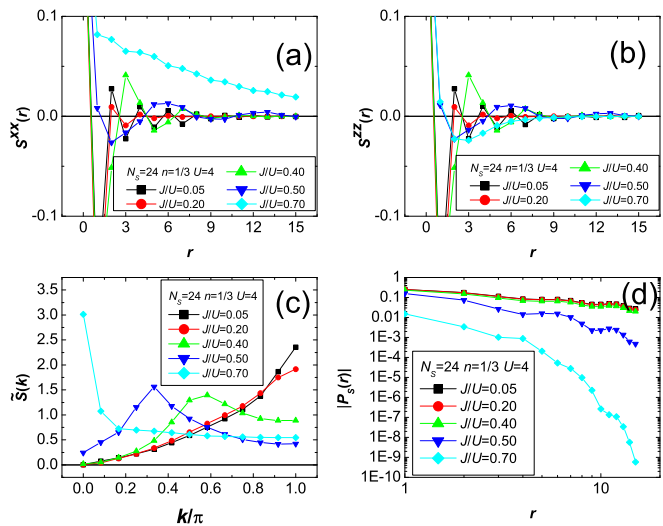}}}
\caption{(Color online) Same as Fig.\ \ref{dataU8}, but for $U=4$.}
\label{dataU4} 
\end{figure}

\begin{figure}
{\centering\resizebox*{8.8cm}{!}{\includegraphics*{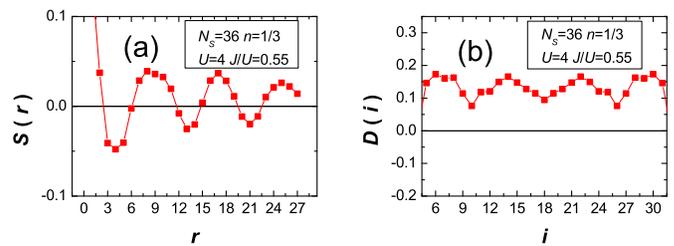}}}
\caption{Results for a lattice with 36 sites, $U=4$, and $J/U=0.55$: (a) the local-moment correlation function $\langle {\bf S}_i\cdot {\bf S}_{i+r} \rangle$, and (b) the dimer order parameter $D(i)$ (see text). 
}
\label{dataU4-SFM} 
\end{figure}

For smaller values of $U$, new magnetic phases appear between SDW and FM, in the intermediate range of $J/U$, such as ilustrated in Fig.\ \ref{dataU4} for $U=4$. 
For $J/U = 0.05$ and 0.20, magnetic correlations remain isotropic with period 2 (SDW), while the system is still superconducting. 
However, for $J/U=0.4$, the system is superconductor, magnetic correlations are still isotropic, but now the peak has shifted to $k=k^* \simeq 0.6\pi$; therefore, in this case, superconductivity coexists with an incommensurate spin-density-wave state (ICSDW) of the local moments. 
Further, when $\tilde{S}(k^*)$ is plotted as a function of $1/N_s$ (not shown), it shows a steady increase with increasing lattice size, which indicates a true quasi--long-range ordered state. 
For $J/U \gtrsim 0.5$, it is useful to consider the correlation function $ S(r)\equiv \langle{\bf S}_i\cdot {\bf S}_{i+r} \rangle$ in conjunction with the dimer order parameter,\cite{Garcia04}
$D(i)=\langle  {\bf S}_i\cdot {\bf S}_{i+1} \rangle,$
which measures the relative orientation of two successive local spins. 
Figure \ref{dataU4-SFM} shows that while $S(r)$ oscillates with $r$, $D(i)$ is always positive, indicating that the local moments are in a spiral state; note also that for $J/U=0.5$ [Fig.\ \ref{dataU4}(c)], $\tilde{S}(0)\neq 0$, which indicates that the local moments are not in a singlet state, thus no longer isotropic. This state is the continuation (to the $U >0$ region) of the  spiral ferromagnetic (SFM) state found for the Kondo lattice model (KLM).\cite{Garcia04}
Figures \ref{dataU4}(c) and (d) respectively show that the FM behavior found for $U=8$ only sets in for $J/U\gtrsim 0.7$, and that 
pairing correlations become strongly suppressed above $J/U\gtrsim 0.5$. 

\begin{figure}
{\centering\resizebox*{8.8cm}{!}{\includegraphics*{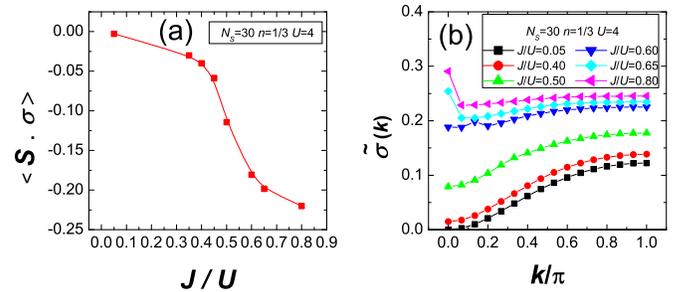}}}
\caption{(Color online) (a) Localized-spin--electron-spin correlation function as a function of $J/U$; (b) magnetic structure factor for the conduction electrons [Eq.\ (\ref{sigmaq})] for different values of $J/U$.}
\label{dataU4-electrons} 
\end{figure}

We now briefly discuss the magnetic behavior of the conduction electrons. 
Figure \ref{dataU4-electrons}(a) shows the correlation between itinerant and local moments on the same site. 
Below $J/U\simeq 0.5$ (the critical value for superconductivity at $U=4$), the conduction electrons are not so strongly correlated with the local moments, as a result of pair formation. 
The magnetic structure factor [defined in a way analogous to that for the local moments, Eq.\ (\ref{sigmaq})] for the conduction electrons [Fig.\ \ref{dataU4-electrons}(b)] shows maxima at $k=\pi$, reflecting the fact that unpaired electrons tend to develop antiferromagnetic-like correlations; nonetheless, 
the range of spatial decay of these magnetic correlations between the itinerant electrons is always much shorter than that of the local moments (data not shown). 
At $J/U=0.5$, both $\tilde{\sigma}(0)$ and $\tilde{S}(0)$ are non-zero, so that rotational symmetry breaks down in each subsystem.  
For $J/U$ sufficiently large, most of the electrons are unpaired, and they follow the magnetic arrangement of the local moments more easily: the magnitude of the local-itinerant correlations increase [Fig.\ \ref{dataU4-electrons}(a)] and the conduction electrons achieve ferromagnetic-like behavior, as evidenced by the peak of $\tilde{\sigma}(k)$ being displaced to $k=0$, as in Fig.\ \ref{dataU4-electrons}(b).  

\begin{figure}[t]
{\centering\resizebox*{7.0cm}{!}{\includegraphics*{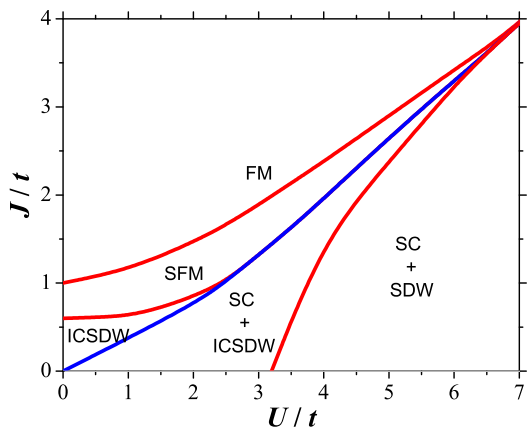}}}
\caption{(Color online) Schematic phase diagram for $n=1/3$: SC stands for superconducting, SDW for spin-density wave, ICSDW for incommensurate SDW, SFM for spiral ferromagnetic and FM for ferromagnetism (see text).}
\label{phase_diagram} 
\end{figure}

Similar analyses have been carried out for other values of $U$, and the results lead to the phase diagram shown in Fig.\ \ref{phase_diagram}, so that the following picture emerges. 
For weak enough Kondo coupling, the paired electrons are superconducting and the unpaired electrons intermediate the coupling between the local moments, leading to the non--pair-breaking SDW state; the latter is commensurate or incommensurate, depending on the strength of $U$, but each subsystem is in a singlet state. 
It is interesting to note that the period-2 SDW is favored in the large-$U$ region; the same happens for other fillings, which indicates that tightly bound pairs can hop more freely (i.e., without being hindered by the Pauli principle) if the local moments are in a period-2 SDW.    
As $J$ increases, more electrons tend to form singlets with the local moments, but superconductivity still survives at the expense of an adjustment of the SDW wavevector. 
At some $J_c(U)$, superconductivity is suppressed: above $U\simeq 2.5$, it is accompanied by the breakdown of spin rotational symmetry; below $U\simeq 2.5$, the suppression of superconductivity takes place within the incommensurate SDW state, and  rotational symmetry breaks down at a larger $J_R(U)$, within the normal phase. 
For $J$ strong enough, the system is normal and ferromagnetic.   

We can now make contact of our results with the $R$Ni$_2$B$_2$C series of compounds, where $R$ is a rare earth. 
Coexistence between superconductivity and some SDW 
is found for $R=$ Tm (ICSDW), Ho and Dy (antiferromagnetism).\cite{Muller01}
Assuming $U$ tracks the Debye temperature, which in turn tracks the inverse ionic radius, we can expect $U$ to grow as $R$ varies from Tm to Dy. 
This trend is consistent with Fig.\ \ref{phase_diagram}: incommensurate SDW's are favored in the small $U$ region of the phase diagram, while commensurate SDW's are favored in the large $U$ region.     

In summary, we have considered a model system composed of local moments coupled through indirect exchange mediated by 
pairing electrons.
The analysis of several quantities calculated for a one-dimensional lattice through DMRG indicates that superconductivity coexists with a magnetically ordered local moment state for a wide range of parameters. 
As the coupling $J$ between the conduction electron and the local moments increases, a superconducting ground state is preserved at the expense of a continuous distortion of the magnetic arrangement, as evidenced by changes in the characteristic wavevector $k^*$.
Superconductivity is suppressed for large enough $J$ by two distinct routes, depending on the range of $U$: 
for large $U$, by the appearance of a pair-breaking magnetic state with broken rotational symmetry 
or, for small $U$, within an ICSDW state.
This model is surely applicable to higher dimensions, preserving most of the qualitative features discussed here. Indeed, we were allowed  
to infer the qualitative trend found in the quaternary borocarbide family of superconductors, in which the magnetic arrangement depends on the rare earth component.

\acknowledgments 
Useful discussions with K. Capelle, D. Garcia, E. Miranda, P. Pagliuso, T.G. Rappoport, and C. Rettori, as well as financial support from the Brazilian Agencies FAPESP, FAPERJ, CNPq,
and FUJB, are gratefully acknowledged.

\bibliography{biblio-kondo-negU}

\end{document}